\documentclass[twocolumn,amsmath,amssymb,prl]{revtex4-1}

\usepackage{graphicx}
\usepackage{dcolumn}
\usepackage{bm}

\begin{document}

\preprint{APS/123-QED}

\title{Nonlinear Hall effect with non-centrosymmetric topological phase in ZrTe$_5$}

\author{Naizhou Wang$^{1}$, Jing-Yang You$^{2}$, Aifeng Wang$^3$, Xiaoyuan Zhou$^3$, Zhaowei Zhang$^1$, Shen Lai$^1$, Hung-Ju Tien$^4$, Tay-Rong Chang$^{4,5,6}$, Yuan-Ping Feng$^{2,7}$, Hsin Lin$^8$, Guoqing Chang$^{1*}$}
\author{Wei-bo Gao$^{1,9}$,}

\email{corresponding author: \\guoqing.chang@ntu.edu.sg, wbgao@ntu.edu.sg}

\affiliation{
$^1$Division of Physics and Applied Physics, School of Physical and Mathematical Sciences, Nanyang Technological University, Singapore 637371, Singapore\\
$^2$Department of Physics, National University of Singapore, 2 Science Drive 3, Singapore 117551, Singapore.\\
$^3$Low Temperature Physics Laboratory, College of Physics and Center for Quantum Materials and Devices, Chongqing University, Chongqing 401331, China\\
$^4$Department of Physics, National Cheng Kung University, Tainan, Taiwan\\
$^5$Center for Quantum Frontiers of Research and Technology (QFort), Tainan, Taiwan.\\
$^6$Physics Division, National Center for Theoretical Sciences, National Taiwan University, Taipei, Taiwan \\
$^7$Centre for Advanced 2D Materials, National University of Singapore, 6 Science Drive 2, Singapore 117546, Singapore \\
$^8$Institute of Physics, Academia Sinica, Taipei 11529, Taiwan \\
$^9$The Photonics Institute and Centre for Disruptive Photonic Technologies, Nanyang Technological University, Singapore 637371, Singapore}

\date{\today}

\begin{abstract}
The non-centrosymmetric topological material has attracted intense attention due to its superior characters as compared to the centrosymmetric one. On one side, the topological phase coming from global geometric properties of the quantum wave function remains unchanged, on the other side, abundant exotic phenomena are predicted to be merely emerged in non-centrosymmetric ones, due to the redistribution of local quantum geometry. Whereas, probing the local quantum geometry in non-centrosymmetric topological material remains challenging. Here, we report a non-centrosymmetric topological phase in ZrTe$_5$, probed by the nonlinear Hall (NLH) effect. The angle-resolved and temperature-dependent NLH measurement reveals the inversion and \emph{ab}-plane mirror symmetries breaking under 30 K, consistent with our theoretical calculation. Our findings identify a new non-centrosymmetric phase of ZrTe$_5$ and provide a platform to probe and control local quantum geometry via crystal symmetries.
\end{abstract}

\maketitle

Since the discovery of the quantum Hall effect, topological states among the condensed matter materials have generated widespread interest owing to their scientific significance and potential for next-generation quantum devices \cite{1,2,3,4,5,6,7,8}. According to whether the material has inversion symmetry, topological materials can be divided into centrosymmetric and non-centrosymmetric \cite{9}. Recently, the non-centrosymmetric topological material attracted intense interest by its non-trivial phenomenon as well as its potential in topological electronics devices. Due to the inversion symmetry break, many exotic properties are predicted to emerge in non-centrosymmetric topological materials, such as topological p-n junction \cite{10}, topological magneto-electric effects \cite{11}, pyroelectricity \cite{12} and surface-dependent topological electronic states \cite{13}. Furthermore, by driving a non-centrosymmetric topological insulator into a superconducting phase, the large upper critical field beyond Pauli limit as well as topological superconductivity with Majorana edge channels can be realized \cite{14,15}. All these non-trivial phenomena make the non-centrosymmetric topological material to be an ideal platform for topological electronics devices and quantum information processing.

Compared with the centrosymmetric topological material, the global topological index is not changed in the non-centrosymmetric one \cite{16,17}. However, due to the difference in local geometric properties of the quantum wavefunction, a series of nonlinear electromagnetic responses, e.g. NLH effects and nonlinear photocurrent, will emerge in non-centrosymmetric topological insulator \cite{18,19,20,21,22,23}. In this work, we predict and discover a new non-centrosymmetric structure ZrTe$_5$ with the space group of \emph{P}\emph{na}2$_1$, which is obtained by slightly translating the ZrTe$_3$ chain along \emph{a}-axis and displacement of Te atoms. The inversion symmetry breaking is observed to emerge below 30 K, leading to NLH effect. The angle-resolved NLH measurement confirms the ab-plane mirror symmetry breaking under 30 K, in consistent with our theoretical prediction. The symmetry breaking is further confirmed by the non-reciprocal transport measurement.

The transition metal pentatelluride ZrTe$_5$ has recently attracted much attention because of its nontrivial topological properties \cite{24,25,26,27,28}. It is predicted that monolayer ZrTe$_5$ is a quantum spin Hall insulator with a large bandgap \cite{24}. ZrTe$_5$ is considered to own an orthorhombic layered structure with the space group of \emph{Cmcm} (D\ensuremath{^{17}}\ensuremath{_{2h}}) \cite{24}. The crystal structure is consisted of the alternate stacking of two-dimensional (2D) layers along the \emph{b}-axis. In each 2D layer, a ZrTe\ensuremath{_{6}} triangular prism forms a one-dimensional ZrTe\ensuremath{_{3}} chain along the \emph{a}-axis. The additional Te ions connect them and consequently form zigzag chains along the \emph{a}-axis, making the crystal tending to grow along the \emph{a}-axis. However, by translating one of the ZrTe$_5$ layers stacked in the \emph{b}-axis of a small amount along the \emph{a}-axis relative to the other layer accompanied by the displacement of Te atoms, we find an energetically more preferred non-centrosymmetric ZrTe$_5$ phase in the space group of \emph{P}\emph{na}2$_1$. Fig. 1(a) shows the structural comparison between the previously reported centrosymmetric ZrTe$_5$ and our predicted non-centrosymmetric one \cite{24}. The average energy of each atom in the non-centrosymmetric one is about 2.7 meV lower than that in the centrosymmetric one. The energy difference between the non-centrosymmetric and centrosymmetric phases indicates the structural phase transition may occur at about 31 K (\ensuremath{\Delta }E/k\ensuremath{_{B}}). The detailed lattice information of centrosymmetric and non-centrosymmetric ZrTe$_5$ are listed in Tables S2 and S3, respectively. The band structure of non-centrosymmetric ZrTe$_5$ was calculated as shown in Fig. 1(b). We found the band splitting (see inset of Fig. 1(b)), which is expected in a system of central inversion symmetry breaking.

\begin{figure}[htp]
\centering
\includegraphics[width=0.48\textwidth]{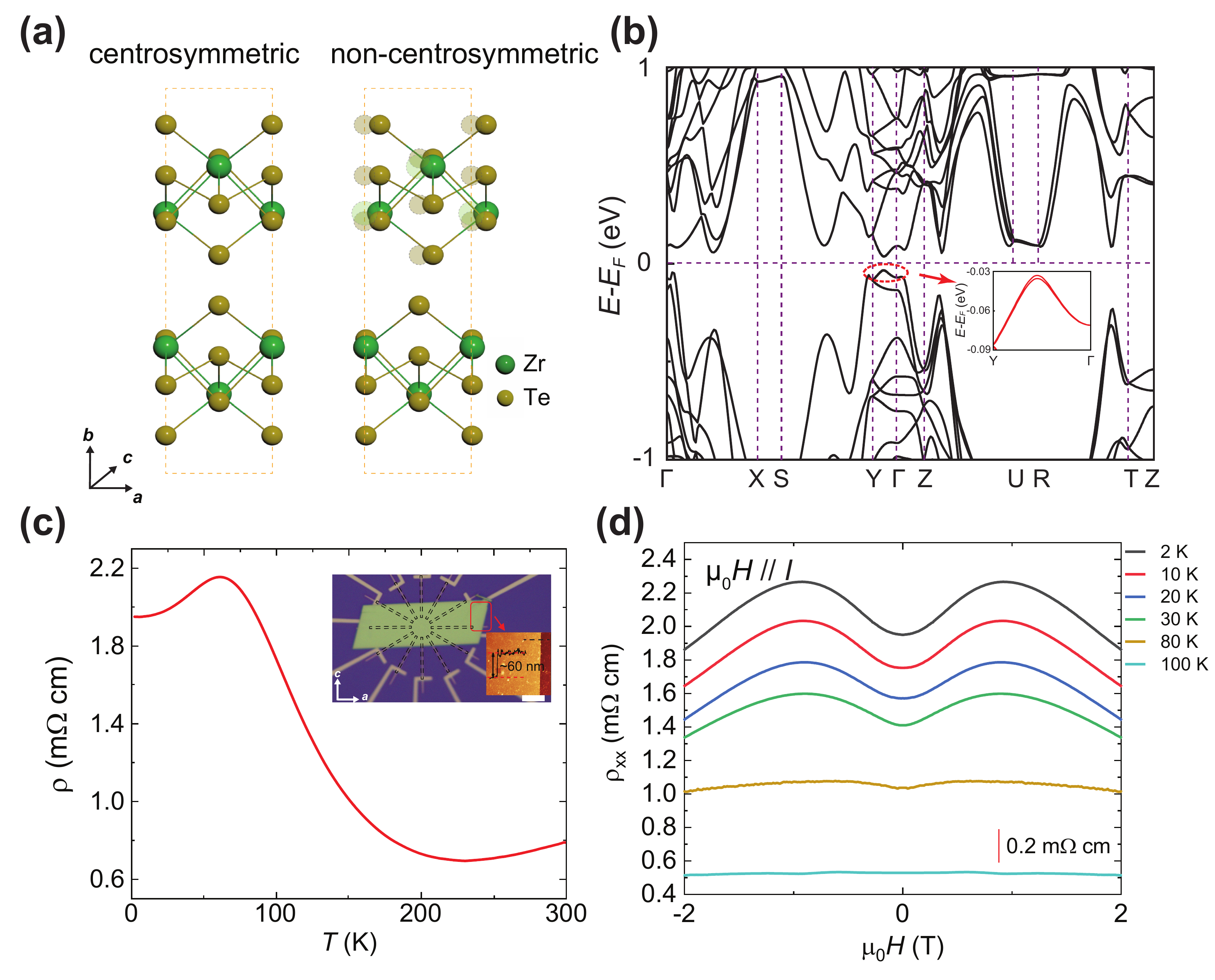}
\caption{The non-centrosymmetric topological phase in ZrTe$_5$ \textbf{(a)} The comparison of crystal structure between centrosymmetric and non-centrosymmetric ZrTe$_5$. For the non-centrosymmetric ZrTe$_5$, the adjacent layer is shift along \emph{a}-axis with a small distance. \textbf{(b)} The calculated band structure of non-centrosymmetric ZrTe$_5$. A valance band splitting could be observed as shown in inset. \textbf{(c)} The temperature dependent resistivity of the circular disc ZrTe$_5$ device. The current is applied along the \emph{a}-axis. The inset shows the optical photo of the device with AFM images. The thickness of the sample is determined to be 60 nm. Scale bar, 10 ${\mu}$m. \textbf{(d)} The magnetic field dependent resistivity the ZrTe$_5$ device, with magnetic field applied along the current direction. A negative magnetoresistance could be observed below 100 K. To make the plots clearer, a vertically shift of 0.2 m${\Omega}$ cm is added.}
\label{Fig. 1}
\end{figure}

In our experiment, high-quality ZrTe\ensuremath{_{5\ }}samples are studied. The circular disc devices with 12 electrodes are used (inset of Fig. 1(c)). Since the ZrTe$_5$ crystal tends to grow along \emph{a}-axis and the exfoliation normally will produce rectangular flakes, we can easily determine the crystal axis and align it to the electrode direction (See supplementary material for details of the sample and device fabrication). Fig. 1(c) shows the temperature dependent resistivity (\emph{RT}) curve of ZrTe$_5$ (60 nm thick) sample when the current is applied along \emph{a}-axis. A resistivity peak is observed at 60 K, which is attributed to slight doping in the thin flake samples (See supplementary material S4) \cite{29}. Negative magnetoresistance under parallel magnetic and electric fields is usually considered as the signature of topological quasiparticles, such as bulk Weyl/Dirac fermions and surface Dirac cones of topological insulators \cite{16,27,30,31,32}. Fig. 1(d) shows the magnetic field dependent resistance for ZrTe$_5$ when the magnetic field is applied along the current direction. A negative magnetoresistance emerges below about 100 K. Our measurements are consistent with our first-principles calculations, where both centrosymmetric and non-centrosymmetric phases of ZrTe$_5$ are topological insulators with protected surface Dirac cones (See supplementary materials Fig. S3). Therefore, we attribute the negative magnetoresistance to be associated with a topological nontrivial order and there is no obvious topological phase change for ZrTe$_5$ below 100 K.

\begin{figure*}[htp]
\centering
\includegraphics[width=0.8\textwidth]{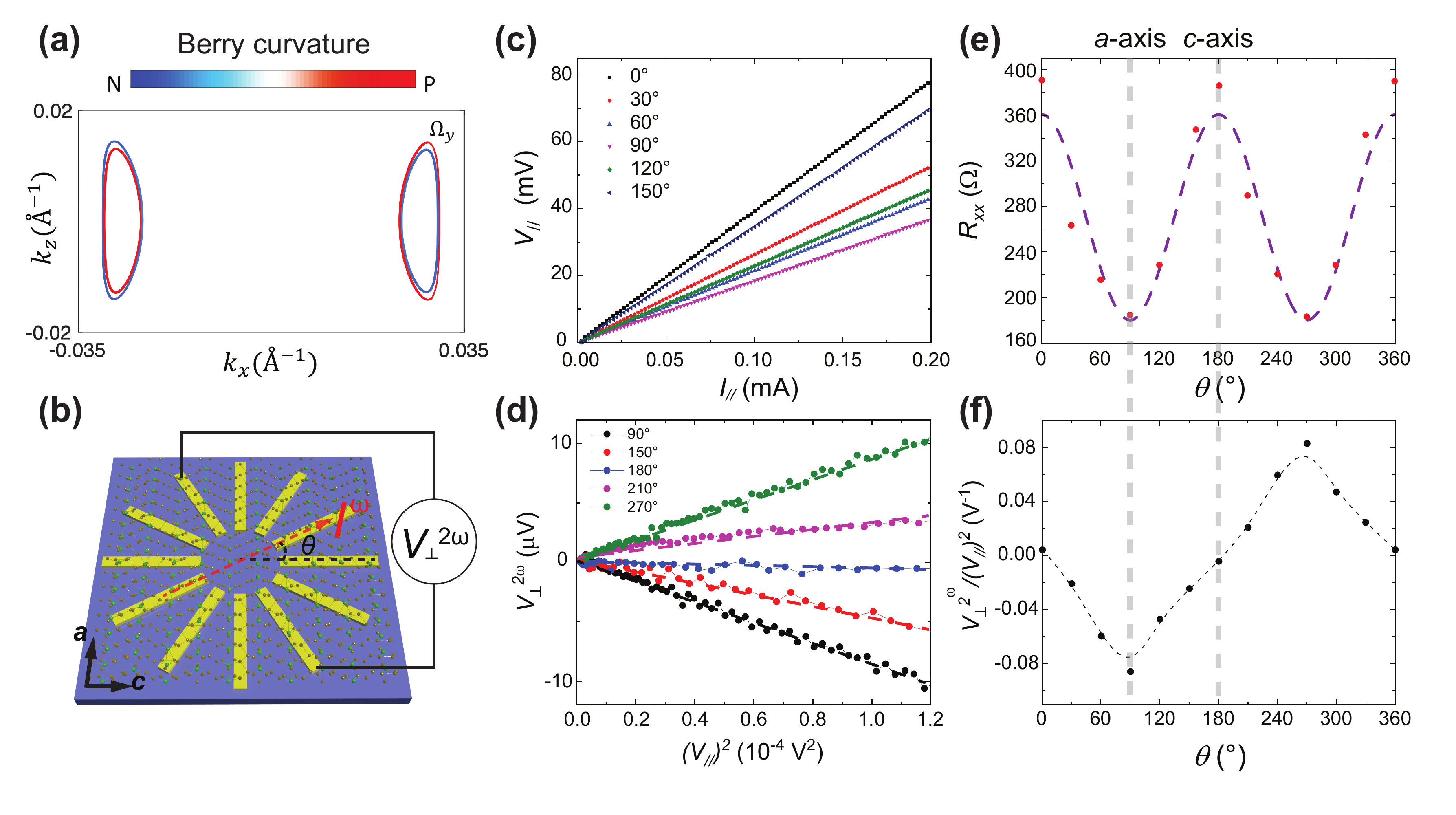}
\caption{The NLH effect in ZrTe$_5$ \textbf{(a)} The Berry curvature distribution calculated based on the non-centrosymmetric ZrTe$_5$. \textbf{(b)} The schematic view of the circular disc ZrTe$_5$ device. The injected current is applied at an angle ${\theta}$ deviated from \emph{c}-axis and the transverse voltage is measured. \textbf{(c)} The first harmonic \emph{I-V} curves for the circular disc ZrTe$_5$ device, with injected current along different direction. \textbf{(d)} The linear dependence of second-harmonic transverse voltage ${V_\perp^{2\omega}}$ on the square of first-harmonic longitudinal voltage ${V_\parallel}$, with injected current along different direction. The round symbols are the experimental data, and the dashed lines are the linear fitting results. \textbf{(e)} The first-harmonic longitudinal resistance as a function of ${\theta}$ in the circular disc ZrTe$_5$ device. ${\theta}$ is the injected current angle measured from \emph{c}-axis. The red solid circle is the experimental data, and the purple dashed line is the fitting result from Eq (2). \textbf{(f)} The second-harmonic Hall response as a function of ${\theta}$ in the circular disc ZrTe$_5$ device. The black solid circles are the experimental data, and the gray dashed line is the fitting result from Eq (3). The error bars are smaller than the symbol. }
\label{Fig. 2}
\end{figure*}

Though the topological phase remains the same in centrosymmetric and non-centrosymmetric ZrTe$_5$, their geometric properties of the wavefunction have changed locally in momentum space. To expose the central inversion symmetry breaking in ZrTe$_5$, we employ the NLH measurement to confirm the redistributed quantum wavefunction: the Berry curvature dipole (\emph{BCD}). The NLH effect has shown its potential to probe crystal symmetry with high sensitivity and accuracy in two-dimensional system such as few layers WTe\ensuremath{_{2}} and twisted WSe\ensuremath{_{2}} \cite{18,19,20,21,22,33}. However, unlike the NLH effect in two-dimension systems, we consider the three-dimensional nature of NLH effect in non-centrosymmetric ZrTe\ensuremath{_{5}}. In a three-dimension system, the NLH current related to the BCD can be represented as\begin{equation} J_a^{2\omega}=\chi_{abc}E_b^\omega E_c^\omega  \end{equation}, where $\chi_{abc} $ is the nonlinear tensor and the E is external electric field \cite{19}. Based on the symmetry analysis of non-centrosymmetric ZrTe$_5$,the only left nonlinear tensor are $\chi_{caa} $ and $\chi_{aac} $. Furthermore, the nonlinear tensor $\chi_{abc} $ is related to the BCD $D_{ab} $. Thus, this NLH current $J_c^{NLHE} $is proportional to $D_{ab} $(See supplementary material for more details of symmetry analysis and derivation of BCD). Fig. 2(a) shows the distribution of Berry curvature contour for certain energy$(\Omega_y) $ in k-space for non-centrosymmetric ZrTe$_5$ crystal structure. Our calculation shows anisotropic Berry curvature contours. As a consequence, the BCD emerges. Hence, when the external electric field is applied along \emph{a}-axis, the NLH response can be measured along the \emph{c}-axis.

\begin{figure}[htp]
\centering
\includegraphics[width=0.48\textwidth]{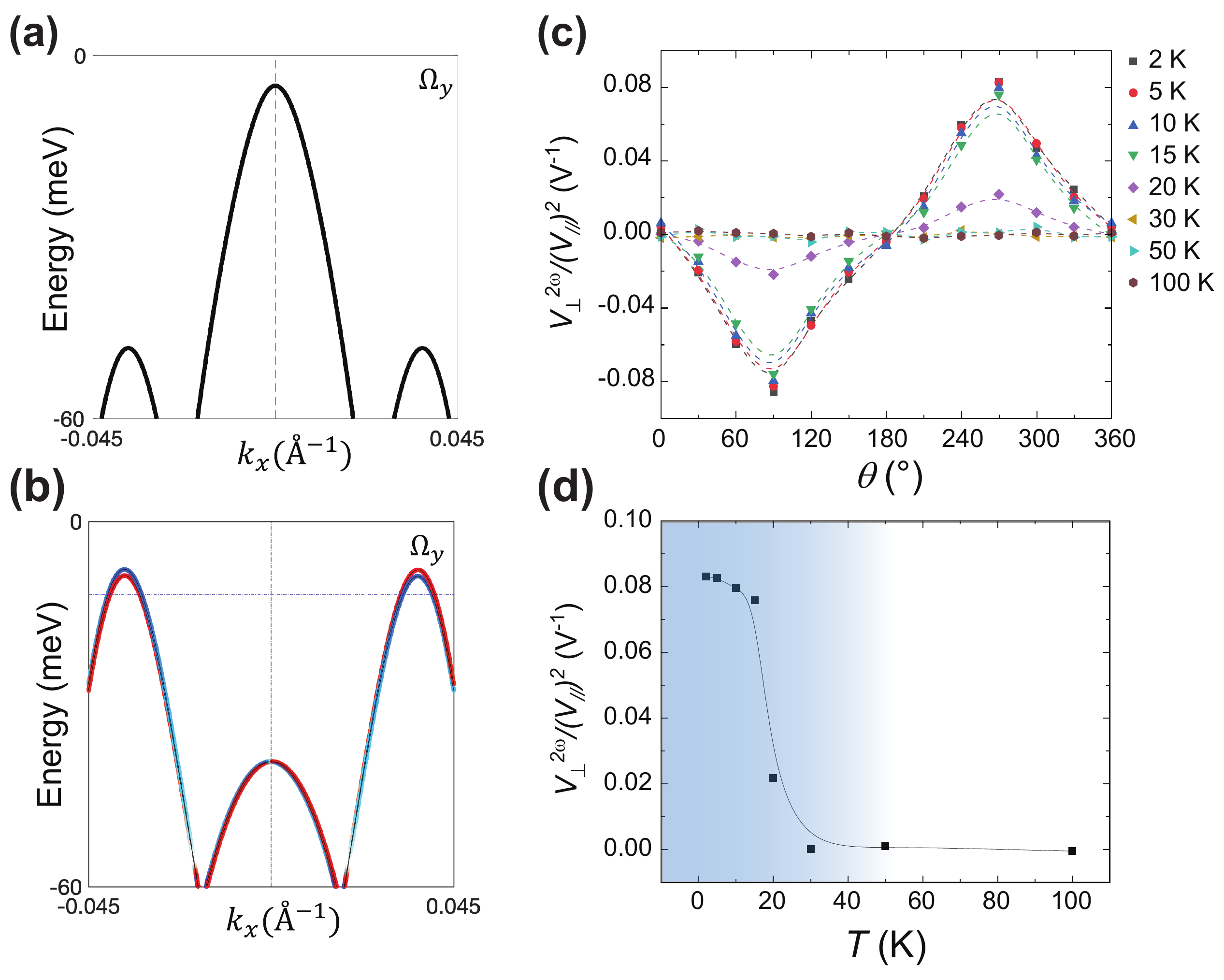}
\caption{The temperature-driven non-centrosymmetric phase transition in ZrTe$_5$.  \textbf{(a)} The distribution of Berry curvature $(\Omega_y) $ for centrosymmetric ZrTe$_5$. The spin-up and spin-down states with opposite Berry curvature overlap with each other due to the spin-degeneracy. \textbf{(b)} The distribution of Berry curvature $(\Omega_y) $ for non-centrosymmetric ZrTe$_5$. Spin-degeneracy is lifted due to the inversion breaking. Thus, a non-zero BCD emerges in the non-centrosymmetric structure. \textbf{(c)} The second-harmonic Hall response as a function of ${\theta}$ at different temperature. The solid symbols are the experimental data, and the dashed lines are the fitting result from eq (3). The second-harmonic Hall response only emerges when temperature lower than 30 K. \textbf{(d)} The temperature dependent second-harmonic Hall response with injected current along \emph{a}-axis. The second-harmonic Hall response suddenly drops above 30 K and remains nearly zero up to 100 K. }
\label{Fig. 3}
\end{figure}

To do the nonlinear transport measurement on the device, an AC current is applied at a fixed frequency (17.777 Hz) along a selected direction of the device. The longitudinal and transverse voltage at both the fundamental and second-harmonic frequencies are measured simultaneously. The angle dependent measurement is carried out with a disk geometry under zero magnetic field at 2 K. The current is injected along one of the 12 electrodes with angle \ensuremath{\theta } specified as the direction deviated from \emph{c}-axis, as shown in Fig. 2(b). To determine the crystal axis as well as the in-plane anisotropy, we analyzed the first-harmonic longitudinal voltage with different current injection directions. As shown in Fig. 2(c), the voltage shows good linear dependence on the injected current at all angles, suggesting excellent ohms contact at each direction. The longitudinal resistance with different angles \ensuremath{\theta }, ${R_\parallel}$ (=${V_\parallel}$/${I_\parallel}$), is shown in Fig. 2(e). The angle dependent longitudinal resistance(${R_\parallel(\ensuremath{\theta })}$) shows a two-fold angular dependence, which is consistent with the ZrTe$_5$ symmetry and previous report\cite{34}. By fitting the curve with formula
 \begin{equation}R_\parallel(\ensuremath{\theta }) = R\ensuremath{_{a}}sin\ensuremath{^{2}}\ensuremath{\theta } + R\ensuremath{_{c}}cos\ensuremath{^{2}}\ensuremath{\theta }  \end{equation}, where R\ensuremath{_{a}} and R\ensuremath{_{c}} denoted as the resistance along \emph{a} and \emph{c} axis, respectively, the in-plane resistance anisotropy coefficient ${\gamma}$ (${\gamma}$= R\ensuremath{_{a}}/R\ensuremath{_{c}}) is obtained as 0.5. We then focus on the second-harmonic part. Indeed, consistent with our predictions, we find that the second-harmonic transverse voltage at the \emph{c}-axis is non-zero when current is along \emph{a}-axis and obeys a linear dependence with the square of current, equivalently, (${V_\parallel}$)$^2$, as shown in Fig. 2(d). Besides, when injected current applied different direction, the second-harmonic voltage varies. When reversing the current direction along \emph{a}-axis (90\ensuremath{^\circ} and 270\ensuremath{^\circ}), the second-harmonic voltage changes its sign, excluding the contribution from sample heating. We have also excluded other possible extrinsic effect to cause the second-harmonic response such as capacitive coupling, contact junctions, flake shape and thermoelectric effect (see supplementary material S7). The slope of ${V_\perp^{2\omega}}$vs. (${V_\parallel}$)$^2$as the function of angle \ensuremath{\theta } is summarized in Fig. 2(f). Unlike the first-harmonic response which shows a two-fold angular dependence (Fig. 2(e)), the second-order response only shows a one-fold dependence. The maximum second-harmonic response is achieved when the current is injected along the \emph{a}-axis and vanish when the current is applied along the \emph{c}-axis.

With the non-centrosymmetric ZrTe$_5$ structure, we further fit the angle resolved nonlinear response through the second-order nonlinear susceptibilities as follows (see supplementary material for detailed derivation)

\begin{equation}
\frac{V_\perp^{2\omega}}{{(V_\parallel)}^{2}}=\rho_csin\theta\bullet\frac{-2{{{cos}^{2}}{\theta d_{15}\gamma^{2}}}+d_{31}{{\gamma^{2}{sin}^{2}}\theta}}{{({cos}^{2}\theta+\gamma{sin}^{2}\theta)}^{2}}
\end{equation}where the d\ensuremath{_{ij}} are the non-vanishing element of the second-order nonlinear susceptibility tensor$\chi_{ijk}^{(2)} $ for the non-centrosymmetric ZrTe\ensuremath{_{5\ }}\cite{35}. Due to the global factor sin\ensuremath{\theta }, it can be inferred that the NLH response is maximal when the driving current is applied perpendicular to the polar \emph{c}-axis. The dashed line in Fig. 2(d) shows the fitting result with above equation, which perfectly capture the experimental data. The observed experimental results ambiguously prove the inversion symmetry breaking in ZrTe$_5$ and perfectly matches with our theoretical prediction.

We further demonstrate the temperature driven phase transition for ZrTe\ensuremath{_{5.}} Figs. 3(a) and 3(b) show the distribution of Berry curvature for centrosymmetric and non-centrosymmetric structures, respectively. The Fermi energy is fixed based on the Shubnikov{\textendash}de Haas (SdH) oscillation results. (See supplementary material S5). Inversion symmetry breaking lifts the spin-degeneracy originally preserved in the centrosymmetric phase (Fig. 3(a)) and creates a Berry curvature pseudo-vector in momentum space (Fig. 3(b)). Our first-principles calculations indicate that the structural phase transition between the non-centrosymmetric and centrosymmetric phases may occur at about 31 K. The NLH effect measurement can probe the redistribution of the quantum wavefunction. To confirm this, we perform the temperature dependent measurements. Fig. 3(c) shows the angle resolved second-harmonic Hall voltage at different temperature, with dashed line showing the fitting result from Eq. (3). The second-harmonic Hall response gradually decreases with increasing temperate. We further plot the temperature dependent second-harmonic Hall response with injected current along \emph{a}-axis in Fig. 2(d). With the increase of temperature, the second-harmonic Hall voltage gradually decreases, suddenly vanishes above 30 K and remains zero up to 100 K. This phenomenon could be reproduced in several devices (see supplementary material S6). The second harmonic Hall response usually follows the scaling behavior with the conductivity, which can be expressed as:

\begin{equation}
\frac{V_\perp^{2\omega}}{{(V_\parallel)}^{2}}\propto\xi\sigma ^{2}+\eta
\end{equation}where \ensuremath{\sigma  } is the conductivity and \ensuremath{\xi} and \ensuremath{\eta } are the constant. For ZrTe$_5$, its conductivity remains nearly unchanged in the temperature range of 2-30 K, while the second harmonic Hall response shows a tremendous change and is different with the description of Eq. (4). The agreement between experimental data and first-principles calculations provides strong evidence for our experimental observation of a temperature driven phase transition from centrosymmetric to non-centrosymmetric in ZrTe$_5$ below 30 K.

\begin{figure*}[htp]
\centering
\includegraphics[width=0.7\textwidth]{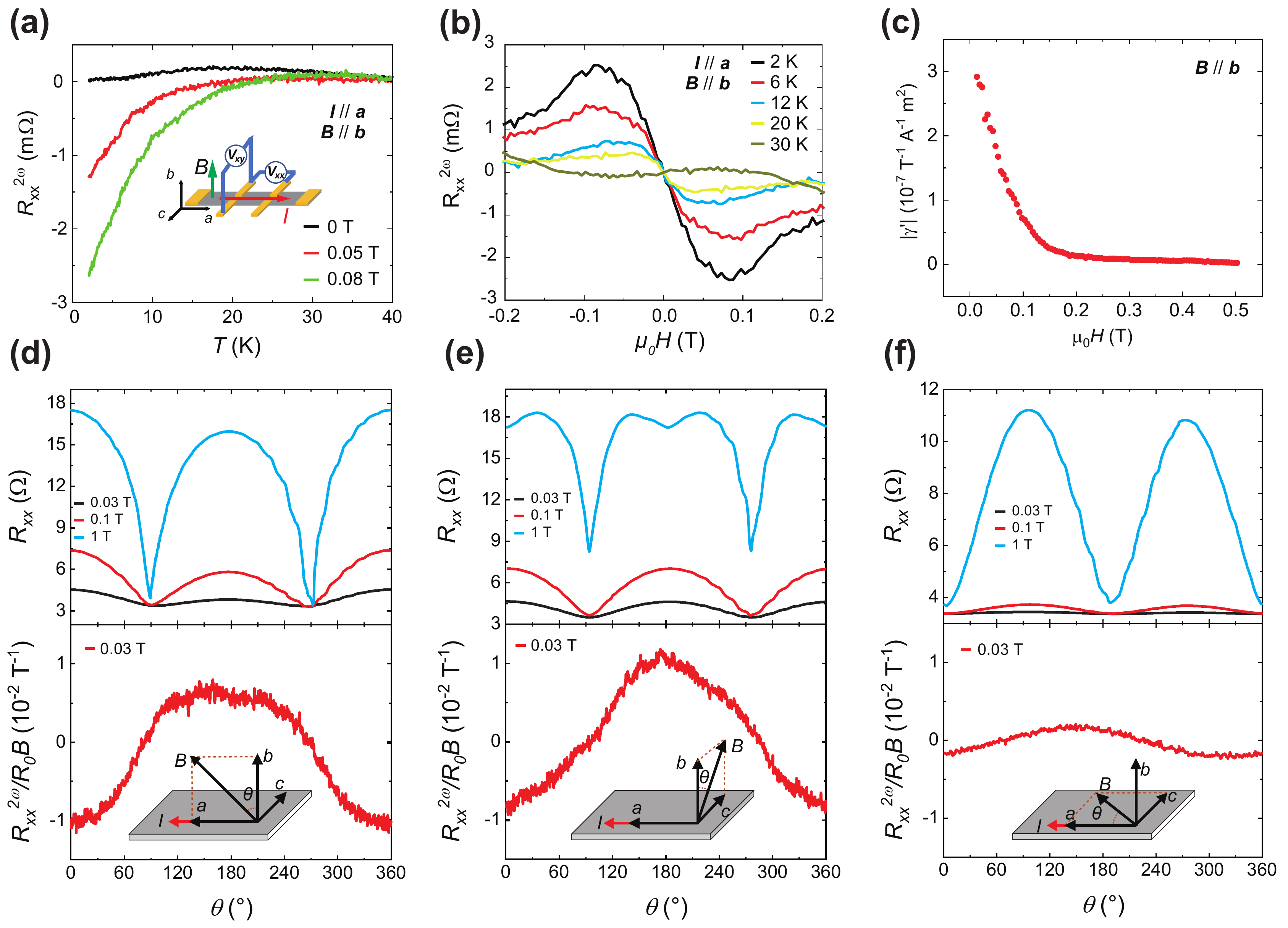}
\caption{The non-reciprocal transport in ZrTe$_5$ \textbf{(a)} The temperature dependent second-harmonic longitudinal resistance of bulk ZrTe$_5$ sample at different magnetic field. The inset shows the schematic view of the Hall bar configuration, where current is applied along \emph{a}-axis and magnetic field along \emph{b}-axis. \textbf{(b)} The magnetic-field dependent second-harmonic longitudinal resistance of bulk ZrTe$_5$ sample at various temperatures. The second-harmonic longitudinal resistance gradually decrease with the increase of temperature and vanishes above 30 K.\textbf{(c)} The magnetochiral anisotropy coefficient of ZrTe$_5$ as a function of magnetic field.  the current is applied along \emph{a}-axis and magnetic field along \emph{b}-axis. \textbf{(d)-(f)} Magnetic-field-orientation dependent resistance and normalized non-reciprocal response ${R_{xx}^{2\omega}/R_0B }$ at different magnetic field, with the magnetic field rotating in the \emph{ab}, \emph{bc}, and \emph{ac} planes, respectively. All data are measured at T = 2 K. The rotation plane and the definition of rotated angle ${\theta}$ are shown in each panel.  }
\label{Fig. 4}
\end{figure*}

In order to exclude the possibility that the phenomenon observed in thin flake ZrTe$_5$ is due to the defect or degradation of sample during fabrication or even interfacial interaction, we perform systematic non-reciprocal transport measurement on a bulk ZrTe$_5$ sample. For an inversion-symmetry broken system, when under magnetic field, a non-reciprocal transport effect could be observed \cite{36}. Depending on specific crystal symmetry, the nonreciprocal resistance could be divided into two types: the chiral structure type with $R=R_0\lbrack1+\gamma(B\bullet I)\rbrack $, where $R_0 $ is the reciprocal resistance and $\gamma $ is a coefficient; and the polar structure with $R=R_0\lbrack1+\gamma(P\times B)\bullet I\rbrack $, where \textbf{P} is the unit vector which shows the direction of polarization in the structure. The coefficient $\gamma $ ($\gamma=\left(\left(\frac R{R_0}\right)-1\right]/(\vert B\vert\bullet\vert I\vert) $), for $B\parallel I~$ with the chiral structure and $(P\times B)\parallel I~$ for the polar structure, could be used to evaluate the magnetochiral anisotropy in the material \cite{36}. Fig. 4(a) shows the second-harmonic longitudinal resistance $R_{xx}^{2\omega} $of bulk ZrTe$_5$ sample under different magnetic field, with current along \emph{a}-axis and magnetic field along the \emph{b}-axis. The second-harmonic resistance $R_{xx}^{2\omega} $ only emerges under 30 K and increases with the increase of magnetic field, which confirms the inversion symmetry breaking in ZrTe$_5$. Moreover, the second-harmonic longitudinal resistance $R_{xx}^{2\omega} $ shows quadratically dependence on the applied current and antisymmetric with the magnetic field, which is in accordance with nonreciprocal response of the polar structure with $R=R_0\lbrack1+\gamma(P\times B)\bullet I\rbrack $ (See supplementary material S8). Fig. 4(b) shows the magnetic field dependent second-harmonic longitudinal resistance $R_{xx}^{2\omega} $ under different temperature. With \ensuremath{\mu }\ensuremath{_{0}}H\textless\ 0.08 T, the $R_{xx}^{2\omega} $ increases with the increase of magnetic field. As increasing temperature, the $R_{xx}^{2\omega} $ gradually decreases and finally vanish at 30 K, which suggests the inversion symmetry breaking happened below 30 K. Our finding in bulk ZrTe$_5$ samples is totally consistent with what is observed in thin flakes sample despite different methods, which proves the centrosymmetric to non-centrosymmetric phase transition in ZrTe$_5$ is intrinsic. Moreover, a giant magnetochiral anisotropy coefficient, ${|\gamma^{'}|=|(2A_\perp R_{2\omega}/(R_0BI_0)| }$, with magnetic field along \emph{b}-axis is observed at low field. We noticed a recent report that also observed a giant magnetochiral anisotropy coefficient as well as the emergence of $R_{xx}^{2\omega} $ below 20 K, which is consistent with our result \cite{37}. Fig. 4(d) to Fig. 4(f) shows the measurement of $R_{xx} $ and $R_{xx}^{2\omega} $ under different magnetic field, with varying different rotating plane along \emph{ab}, \emph{bc} and \emph{ac} planes, respectively. For $R_{xx} $, the resistance shows a two-fold dependency on the angle \ensuremath{\theta } within \emph{ab}, \emph{bc} and \emph{ac} plane. As depicted from the equation that $R=R_0\lbrack1+\gamma(P\times B)\bullet I\rbrack $, we can determine the polar axis by measuring $R_{xx}^{2\omega} $ with magnetic field rotating along different crystalline plane. In order to prevent the influence of reciprocal response, $R_{xx}^{2\omega} $ is normalized by $R_0B $. For \emph{ab}-plane and \emph{bc}-plane, the $R_{xx}^{2\omega} $ follows a cos\ensuremath{\theta } dependence, where \ensuremath{\theta } is the angle that magnetic field deviates from \emph{b}-axis. In contrast, the $R_{xx}^{2\omega} $ remains almost zero when rotating magnetic within \emph{ac}-plane. Since the current is along \emph{a}-axis, it suggests that the polar axis \textbf{P} is along the \emph{c}-axis, in accordance with the results from NLH measurement and first-principal calculation.

In summary, our work highlights a phase transition from centrosymmetric to non-centrosymmetric structure at 30 K in ZrTe$_5$. The symmetry breaking leads to local quantum geometry redistribution with global geometry unchanged. The phase transition is probed by temperature dependent, angle resolved 3D NLH effect, as well as non-reciprocal transport measurement. Furthermore, our results provide a valuable tool to identify and control the local quantum geometry via crystal symmetries, which could be extended to a broad range of topological materials. Finally, it paves the way to realize more interesting applications such as signal rectification or frequency doubling, topological p-n junction \cite{10} and topological magneto-electric effects \cite{11}.

\begin{acknowledgments}
N.Z.W., Z.Z., S.L. and W.B.G. thank the financial support from the Singapore National Research Foundation through its Competitive Research Program (CRP Award No. NRF-CRP21-2018-0007, NRF-CRP22-2019-0004), QEP programme, Singapore Ministry of Education (MOE2016-T3-1-006 (S)). G.C. acknowledges the support of the National Research Foundation, Singapore under its Fellowship Award (NRF-NRFF13-2021-0010) and the Nanyang Assistant Professorship grant from Nanyang Technological University. J.Y.Y. and Y.P.F. is supported by the Ministry of Education, Singapore, under its MOE AcRF Tier 3 Award MOE2018-T3-1-002. A.F.W and X.Y.Z. is financially supported by the National Natural Science Foundation of China (Grants No. 12004056, No. 52071041). H.L. acknowledges the support of the Ministry of Science and Technology (MOST) in Taiwan under grant number MOST 109-2112-M-001-014-MY3. H.J.T. and T.-R.C. is supported by the Young Scholar Fellowship Program from the Ministry of Science and Technology (MOST) in Taiwan, under a MOST grant for the Columbus Program MOST110-2636-M-006-016, the National Cheng Kung University, Taiwan, and National Center for Theoretical Sciences, Taiwan. Work at NCKU is supported by MOST, Taiwan, under grant MOST107-2627-E-006-001 and Higher Education Sprout Project, Ministry of Education to the Headquarters of University Advancement at NCKU.
\end{acknowledgments}

\vspace{12 pt}N.Z.W. and J.Y.Y. contribute equally to this work.

\end{document}